\documentclass[11pt,a4paper,twocolumn]{article}

\usepackage{authblk}
\usepackage{amssymb}
\usepackage{amsmath}
\usepackage{amsfonts}
\usepackage{latexsym}
\usepackage[dvips]{epsfig}
\usepackage{graphicx}

\usepackage[numbers]{natbib}

\usepackage{fullpage}
\usepackage{times}
\usepackage{upgreek}
\usepackage{color}
\usepackage{bm}
\usepackage{cuted}

\def \curl{\mbox{curl\hskip 1pt}}
\def \Curl{\mbox{Curl\hskip 1pt}}
\def \div{\mbox{div\hskip 1pt}}
\def \Div{\mbox{Div\hskip 1pt}}
\def \tr{\mbox{tr\hskip 1pt}}

\renewcommand{\vec}[1]{\boldsymbol{#1}}

\renewcommand{\ddot}{\boldsymbol{\cdot}}

\title{Electro-elastic Lamb waves in dielectric plates}
\author[1]{Hannah Conroy Broderick}
\author[2]{Luis Dorfmann}
\author[1]{Michel Destrade}

\affil[1]{School of Mathematics, Statistics and Applied Mathematics, NUI Galway,  \newline University Road, Galway, Ireland} 
\affil[2]{Department of Civil and Environmental Engineering, Tufts University, \newline Medford, MA 02155, USA}

\begin{document}

\maketitle
  
\begin{abstract}

We study the propagation of Lamb waves in soft dielectric plates subject to mechanical and electrical loadings. 
We find explicit expressions for the dispersion equations in the cases of neo-Hookean and Gent dielectrics.
We elucidate the effects of the electric field, of the thickness-to-wavelength ratio, of pre-stress and of  strain-stiffening  on the wave characteristics.

\end{abstract}

\section{Introduction} 

Wave propagation in soft dielectric materials has been shown to depend on both the underlying deformation and the applied electric field for a variety of deformations and geometries.

Dorfmann and Ogden \cite{DorfOg10} derived the incremental formulation and showed the dependence of the surface wave velocity on the electric field in an electro-elastic half-space (voltage-controlled case). 
 The case of a dielectric plate under plane strain was investigated by Shmuel et al. \cite{Shmuel12}, who showed the effects of pre-stretch and applied electric displacement on the wave velocity (charge-controlled case). 
 There has been significant work on cylinders and tubes \cite{ChenDai12, Shmuel13, Su16, Wu17}, showing that the velocity of the wave depends on the electric field and the direction in which it is applied, as well as the direction of propagation relative to the underlying electro-mechanical deformation. In particular, Wu et al. \cite{Wu17} highlighted the possibility of using wave propagation to detect defects or cracks in the material based on their analysis of circumferential waves.

The dependence of the wave velocity on the applied electric field suggests the possibility of controlling the velocity of the propagating wave by applying an appropriate electric field \cite{Shmuel12}. 
It also paves the way for applying acoustic non-destructive evaluation techniques to dielectric  plates.

Here, we first recall the equations governing the large deformation of a soft dielectric plate and then the subsequent propagation of small-amplitude Lamb waves, using incremental theory (Section \ref{section2}). 
We write the equations of motion in the Stroh form, and then solve them for the neo-Hookean dielectric model (Section \ref{solution}).
We separate symmetric from antisymmetric modes of propagation to obtain explicit expressions for the corresponding dispersion equations, and for their limits in the long wavelength-thin plate and short wavelength-thick plate regimes.
In Section \ref{results}, we solve the dispersion equations numerically to highlight the effects of loading and geometry on the wave propagation.
We also use the Gent dielectric model to look at the effects of strain-stiffening and snap-through on the plate's acoustics.


\section{Equations of motion} 
\label{section2}


We use the incremental theory of electro-elasticity \cite{DorfOg10} to analyse  wave propagation in a finitely deformed electro-active plate. 
For completeness of presentation we  first summarise the main parts of the theory.  

The incremental (small-amplitude) mechanical  displacement is denoted $\mathbf{u}$ and the increments of the total nominal stress and of the Lagrangian forms of the electric displacement and the electric field are denoted by  $\mathbf {\dot T},  \mathbf{\dot D}_{\mathrm L}, \mathbf{\dot E}_{\mathrm L}$, respectively. For an incompressible electro-elastic material  the corresponding push-forward measures are obtained as 
\begin{equation}
\label{push-forward}
\mathbf {\dot T}_0=\mathbf F  \mathbf {\dot T},\quad \mathbf {\dot D}_{\mathrm L0}=\mathbf F \mathbf{\dot D}_{\mathrm L},\quad \mathbf {\dot E}_{\mathrm L 0}=\mathbf F^{-\mathrm T} \mathbf{\dot E}_{\mathrm L},
\end{equation} where $\mathbf F$ is the deformation gradient.  
The incremental quantities satisfy the Lagrangian form of the governing equations
\begin{equation}
\label{inc-equations}
\Div \mathbf {\dot T}=\rho  \mathbf u_{,tt},\quad  \Div \mathbf{\dot D}_{\mathrm L}=0,\quad \Curl \mathbf {\dot E}_{\mathrm L}=\mathbf 0,
\end{equation}
and, equivalently,  their updated (Eulerian) forms
\begin{equation}
\label{up-inc-equations}
\div \mathbf {\dot T}_0=\rho \mathbf u_{,tt},\quad  \div \mathbf{\dot D}_{\mathrm L 0}=0,\quad\curl \mathbf {\dot E}_{\mathrm L 0}=\mathbf 0,
\end{equation}
where $\rho$ is the (constant) mass density per unit volume  and $,t$ denotes the time derivative.
 
With no external field and  no applied mechanical traction the incremental boundary conditions have the simple forms
\begin{equation}
\label{inc-BC}
\dot {\mathbf T}^{\mathrm T} \mathbf N=\mathbf 0, \quad \mathbf {\dot E}_{\mathrm L }\times \mathbf N=\mathbf 0,\quad \mathbf {\dot D}_{\mathrm L }\ddot \mathbf N=\dot \sigma_{\mathrm F},
\end{equation}
where $\dot \sigma_{\mathrm F}$ identifies an increment in the referential charge density on the electrodes attached to the major surfaces of the plate.

Superposed on the current configuration we consider an incremental motion, tracked by the incremental deformation gradient $\mathbf {\dot F}$, combined with an increment in the electric field $\mathbf{\dot E}_{\mathrm L}$. This results in  increments of the total nominal stress  and of the Lagrangian electric displacement field as specified by the incremental forms of the constitutive equations \cite{DorfmannOgden2019}. 

In particular, for an incompressible material we have $\det\mathbf F=1$ at all times. Then, $\mathbf{\dot T}$ and $\mathbf{\dot D}_{\mathrm L}$  have the forms
\begin{equation}
\label{inc-stress}
\mathbf {\dot T}=\boldsymbol{\mathcal A} \mathbf {\dot F} + \boldsymbol{\mathbb A} \mathbf {\dot E}_{\mathrm L}-\dot p \mathbf F^{-1}+p \mathbf F^{-1}\mathbf {\dot F} \mathbf F^{-1},
\end{equation} and
\begin{equation}
\label{inc-D_L}
\mathbf {\dot D}_{\mathrm L}=-\boldsymbol{\mathbb A}^{\mathrm T} \mathbf {\dot F}- \boldsymbol{\mathsf A} \mathbf {\dot E}_{\mathrm L},
\end{equation}
where $\boldsymbol{\mathcal A}$, $\boldsymbol{\mathbb A}$, $\boldsymbol{\mathsf A}$ are, respectively, the fourth-, third- and second-order electro-elastic moduli tensors, and $p$ is a Lagrange multiplier due to incompressibility (and $\dot p$ is its increment).   
The use of  \eqref{push-forward} gives the updated forms of the  incremental constitutive equations 
\begin{equation}
\label{up-inc-stress}
\mathbf{\dot{T}}_0=\boldsymbol{\mathcal{A}}_0\mathbf{L}+\boldsymbol{\mathbb{A}}_0\mathbf{\dot{E}}_{\mathrm{L}0}
+p\mathbf{L}-\dot{p}\mathbf{I},
\end{equation}
and
\begin{equation}
\label{up-inc-D}
 \mathbf{\dot{D}}_{\mathrm{L}0}=-\boldsymbol{\mathbb{A}}_0^\mathrm{T}\mathbf{L} -\boldsymbol{\mathsf{A}}_0\mathbf{\dot{E}}_{\mathrm{L}0},
\end{equation}
where  $\mathbf I$ is the identity tensor and  $\mathbf L$  the gradient of the  Eulerian version of the incremental displacement vector $\mathbf u$, see \cite{Ogden2014} for details. The latter satisfies the incremental incompressibility condition
\begin{equation}
\label{inc-constraint}
\tr \mathbf L = \div\mathbf{u}=0.
\end{equation}

From \eqref{inc-BC} we find the updated incremental boundary conditions 
\begin{equation}
\label{up-inc-BC}
\mathbf{\dot T} _0^{\mathrm T} \mathbf n=\mathbf 0, \quad \mathbf {\dot E}_{\mathrm L 0 }\times \mathbf n=\mathbf 0,\quad \mathbf {\dot D}_{\mathrm L 0}\cdot \mathbf n=\dot \sigma_{\mathrm F 0},
\end{equation}
where $\dot {\sigma}_{\mathrm F0}$ is the Eulerian form of the charge density increment.  

In what follows we focus attention on isotropic and incompressible electro-elastic materials with properties dependent on just two invariants, denoted $I_1$ and $I_5$, and defined as
\begin{equation}
\label{I1I5}
I_1=\tr \mathbf C, \quad I_5=\mathbf E_{\mathrm L} \ddot \mathbf C^{-1}\mathbf E_{\mathrm L},
\end{equation} where $\mathbf C=\mathbf F^{\mathrm T}\mathbf F$ is the right Cauchy-Green deformation tensor.

We denote the  lateral dimensions and the total thickness of an electroelastic plate in the undeformed configuration by $2 L$ and $ 2 H$, respectively.
The deformed configuration is defined using the Cartesian coordinates  $(x_1, x_2, x_3)$ with $x_2$ oriented normal to the major surfaces. We focus on equi-biaxial deformations with principal stretches  
\begin{equation}
\label{stretches}
\lambda_1=\lambda_3=\lambda,\quad \lambda_2=\lambda^{-2}. 
\end{equation}
The deformed plate is then in the region
\begin{equation}
\label{DefGeometry1}
-  \ell \le x_1\le  \ell, \quad -h \le x_2\le h,\quad  - \ell \le x_3\le  \ell,
\end{equation}
where $2 h =2 \lambda^{-2}  H$ and $ 2\ell = 2\lambda\, L$ are the plate's dimensions in the deformed configuration.

In addition,  a potential difference (voltage) is applied between the compliant electrodes attached at the top and bottom surfaces. The in-plane dimensions are much larger than the thickness and the edge effects can therefore be neglected. It follows that the accompanying electric  and  electric displacement fields have a single component each, denoted $E_2$ and $D_2$, respectively, along the normal to the major surfaces of the plate.  Equation \eqref{push-forward}, specialised to the underlying configuration gives the Lagrangian counterparts as
\begin{equation}
\label{Lag-Fields}
E_{\mathrm L2}=\lambda^{-2} E_2,\quad D_{\mathrm L2}=\lambda^2 D_2.
\end{equation} while the invariants \eqref{I1I5} have the simple forms
\begin{equation}
\label{I1I5again}
I_1=2 \lambda^2+\lambda^{-4},\quad I_5=\lambda^4E_{\mathrm L2}^2.
\end{equation}

Superimposed on the underlying configuration we  consider two-dimensional electro-elastic increments with components $u_1,u_2, \dot E_{\mathrm L01},\dot E_{\mathrm L02}$, which depend on $x_1,x_2$ and $t$ only. 

Equation \eqref{up-inc-equations}$_3$ suggests the introduction of a scalar function $\varphi=\varphi(x_1,x_2)$ such that 
\begin{equation}
\label{phi}
\dot E_{\mathrm L01}=-\varphi_{,1},\quad \dot E_{\mathrm L02}=-\varphi_{,2},
\end{equation}
where the subscripts 1 and 2 following a comma denote partial derivatives with respect to $x_1$ and $x_2$, respectively. 
Using \eqref{up-inc-stress} we find the non-zero incremental stress components $\dot T_{011}, \dot T_{012}, \dot T_{021}$, and $ \dot T_{022}$ and the non-zero incremental electric displacement components $\dot D_{\mathrm L01}$ and $\dot D_{\mathrm L02}$ from \eqref{up-inc-D} \cite{DorfOg10, Ogden2014, Su18}. Equation \eqref{up-inc-equations}$_1$ and \eqref{up-inc-equations}$_{2}$ then specialise to
\begin{align}
&\dot T_{011,1}+\dot T_{021,2} =\rho u_{1,tt},\nonumber\\
&\dot T_{012,1}+\dot T_{022,2} =\rho u_{2,tt},\label{exp-equilibrium}\\
&\dot D_{\mathrm L01,1}+\dot D_{\mathrm L02,2} =0,\label{divD}
\end{align}
together with the incompressibility condition \eqref{inc-constraint}  
\begin{equation}
\label{divu2}
u_{1,1}+u_{2,2}=0.
\end{equation} 

It remains to specify the incremental boundary conditions \eqref{up-inc-BC}$_{1,2}$  on the major surfaces $x_2=\pm h$, as
\begin{equation}
\label{exp-BC}
\dot T_{021}=\dot T_{022}=0,\quad \dot E_{\mathrm L01}=0,
\end{equation}
while the boundary condition \eqref{up-inc-BC}$_3$ is not used.

Specifically, we seek solutions with sinusoidal dependence in the $x_1$ direction for the variables $u_1,u_2,\dot D_{\mathrm L02}, \dot T_{021},\dot T_{022},\varphi$ in the form
\begin{align}
&\left\{u_1,u_2,\dot D_{\mathrm L02}, \dot T_{021},\dot T_{022},\varphi\right\}\label{StrohVector}\\
&=\Re \left\{\left[ U_1, U_2, \textrm i k \Delta , \textrm i k \Sigma_{21}, \textrm i k \Sigma_{22}, \Phi \right]  e^{\textrm i k(x_1 - vt)} \right \}\nonumber,
\end{align}
where the constant $k$ is the wave number, $v$ the wave speed and  the amplitude functions $U_1$, $U_2$, $\textrm i k \Delta$, $\textrm i k \Sigma_{21}$, $\textrm i k \Sigma_{22}$, $\Phi$ are functions of $k x_2$ only. 

In what follows, it is convenient to consider the variables $ U_1$, $U_2$, $\Delta$ as the components of a generalised displacement vector $\mathbf U$, and the variables $\Sigma_{21}$, $\Sigma_{22}$, $\Phi$  as the components of a generalised traction vector $\mathbf S$.  
We find that the governing equations can then be arranged in the Stroh form 
\begin{equation}
\label{StrohForm0}
\boldsymbol \eta' = \mathrm i \mathbf N \boldsymbol \eta,
\end{equation} 
where $\boldsymbol \eta=\left(\mathbf U,\mathbf S\right)^{\mathrm T}$ is the Stroh vector and the prime denotes differentiation with respect to $kx_2$. The $6 \times 6$ matrix $\mathbf N$ has the form
\begin{equation}
\label{StrohForm}
\mathbf{N} = \left[ 
\begin{array}{cc} 	
\mathbf{N}_1 & \mathbf{N}_2 \\
\mathbf{N}_3 & \mathbf{N}_1^{\mathrm T}
\end{array} \right],
\end{equation}
where the $\mathbf N_i$ are $3\times 3$ sub-matrices and $\mathbf N_2,\mathbf N_3$  are symmetric \cite{Shuv00, Su18, DorfOg19}. 

Next, we introduce shorthand notations for the coefficients in  \eqref{up-inc-stress} and \eqref{up-inc-D}. For an energy function $\Omega$ that depends linearly on the invariants $I_1$ and $I_5$, we use
\begin{align}
 a &=\mathcal A_{01212}=2 \left(\lambda^2 \Omega_1 + \lambda^2 E_{L2}^2 \Omega_5 \right),\nonumber\\
 c &=\mathcal A_{02121}=2  \lambda^{-4} \Omega_1 ,\nonumber\\
2b &=\mathcal A_{01111}+\mathcal A_{02222}-2\mathcal A_{01122}-2\mathcal A_{01221}\nonumber\\
&=4(\lambda^2-\lambda^{-4})^2\Omega_{11}+a+c,\nonumber\\
d &=\mathbb A_{0211}=-2 \lambda^2 E_{L2} \Omega_5, \notag\\
e &=\mathbb A_{0222}-\mathbb A_{0112}=2d,\nonumber\\
f &=\mathsf A_{011}=\mathsf A_{022}=2\Omega_5, \label{coefficients}
\end{align}
where $\Omega_j = \partial \Omega / \partial I_j$ for $j=1,5$. We also recall the connections \cite{DorfmannOgden2019}
\begin{equation}
\nonumber
\mathcal A_{0jilk}-\mathcal A_{0ijlk}=\left(\tau_{jl}+p\delta_{jl}\right)\delta_{ik}-\left(\tau_{il}+p\delta_{il}\right)\delta_{jk},
\end{equation}
which for $\tau_{22}=0$ results in $p=c$.

We introduce  dimensionless versions of \eqref{coefficients} as follows 
\begin{align}
\bar a &= a/\mu, & \bar b &= b/\mu, & \bar c &= c/\mu, \notag \\
\bar d &= d/\sqrt{\mu\varepsilon}, & \bar e &= e/\sqrt{\mu\varepsilon}, & \bar f &= f/\varepsilon, \label{coeff-nd}
\end{align} where  $\mu$ is the initial shear modulus associated with  purely elastic deformations and $\varepsilon$ is the (constant) electric permittivity. Together with the dimensionless field measures
\begin{align}
\bar E_0 &= E_{L2} \sqrt{\varepsilon/\mu}, & \bar D_0 = D_{L2}/\sqrt{\mu\varepsilon}, 
\end{align} 
and the non-dimensional components of the Stroh vector $\vec{\eta}$, 
\begin{align}
\bar U_i &= U_i, & \bar \Delta &= \Delta/\sqrt{\mu\varepsilon}, \notag \\
\bar \Sigma _{2i} &= \Sigma_{2i}/\mu, & \bar \Phi &= \Phi\sqrt{\varepsilon/\mu}, \label{eta-nd}
\end{align} 
we arrive at a non-dimensional version of the Stroh formulation \eqref{StrohForm0} \cite{Su18}.

To derive the Stroh equations, we follow the procedure in \cite{Su18}. 
As the only change in the governing equations from \cite{Su18} is to the equilibrium equation for the stress, i.e. \eqref{up-inc-equations}$_1$ or equivalently \eqref{exp-equilibrium}, the only entries that change are those related to $\bar \Sigma_{21}$ and $\bar \Sigma_{22}$. 
As a result, $\mathbf{N}_1$ and $\mathbf{N}_2$ are identical to the forms derived in \cite{Su18}, and $\mathbf{N}_3$ becomes
\begin{equation}
\mathbf{N}_3 =  \begin{bmatrix}  \dfrac{\bar e^2}{\bar f}-2(\bar b+\bar c)+ \bar v^2 & 0 & -\dfrac{\bar e}{\bar f} \\[12pt]
0 & \bar c- \bar a +\bar v^2 & 0 \\[12pt]
-\dfrac{\bar e}{\bar f} & 0 & \dfrac{1}{\bar f} \end{bmatrix},
\end{equation} where $\bar v^2 = \rho v^2 / \mu$ is a non-dimensional version of $v^2$.


\section{Resolution for the neo-Hookean dielectric plate } 
\label{solution}


We consider solutions in the form  $\boldsymbol \eta= \boldsymbol \eta_0 \mathrm e^{-p k x_2}$, which reduce  \eqref{StrohForm0}   to an eigen-problem. Hence,  the eigenvalues and eigenvectors $p_{j}, \boldsymbol \eta^{(j)}, j=1,\dots,6$ are determined by solving the characteristic equation
\begin{equation}
\label{eigen-problem}
\det \left(\mathbf N-\mathrm i p\boldsymbol {\mathbf I}\right)\boldsymbol \eta_0= 0,
\end{equation}
where $\boldsymbol {\mathbf I}$ is the $6 \times 6$ identity matrix. 
Note that the form of the solution used here is equivalent to the one  in \cite{Su18} using $p=\mathrm i q$.  
It follows that  the general solution for $\boldsymbol \eta$ has the form
\begin{equation}
\label{g-solution}
\boldsymbol \eta= \sum_{j=1}^6 c_j \boldsymbol \eta^{(j)}\mathrm e^{- p_j k x_2},
\end{equation}
where $c_j, j=1,\dots,6$ are constants to be determined by the boundary conditions \eqref{exp-BC}. 

To illustrate the solution, we now specialise the constitutive model $\Omega(I_1,I_5)$ to the neo-Hookean electroelastic form \cite{ZhSu07}
\begin{equation}
\label{neo-Hookean}
\Omega_{\mathrm{nH}}=\frac{\mu}{2}(I_1-3)-\frac{\varepsilon}{2}I_5.
\end{equation}
Then we find that solving the characteristic equation \eqref{eigen-problem} results in
\begin{align}
p_1 &= -p_4=1, \qquad  p_2 =-p_5=1,\notag \\
p_3 &=-p_6=\lambda^2 \sqrt{\lambda^2-\bar v^2}, \label{eigenvalues}
\end{align}
with corresponding eigenvectors 
\begin{align}
&\boldsymbol \eta^{(1)}= \left[
\begin{array}{c}
\lambda^4 \\
\mathrm i \lambda^4 \\
2\lambda^6 \bar E_0\\
2 \mathrm i\\
\lambda^4\left(\bar v^2-\lambda^2+\lambda^4\bar E_0\right)-1\\
0
\end{array}
\right], 
\notag \\
&\boldsymbol \eta^{(2)}=\left[
\begin{array}{c}
\mathrm i \lambda^6 \bar E_0^2\\
- \lambda^6 \bar E_0^2 \\
\mathrm i\left(\lambda^6- \lambda^4 \bar v^2+\lambda^8\bar E_0^2+ 1\right)\\
\lambda^2\bar E_0\left(\lambda^6-\lambda^4\bar v^2 -\lambda^{8}\bar E_0^2-1\right)\\
0\\
\lambda^4\left(\bar v^2-\lambda^2+\lambda^4\bar E_0^2\right)-1
\end{array}
\right],
\notag 
\\
&\boldsymbol \eta^{(3)}= \left[
\begin{array}{c}
\mathrm i \lambda^6 \sqrt{\lambda^2-\bar v^2} \\
- \lambda^4 \\
\mathrm i \lambda^8\bar E_0\sqrt{\lambda^2-\bar v^2}\\
\lambda^4\left(\bar v^2-\lambda^2-\lambda^4\bar E_0^2\right)-1\\
-2\mathrm i \lambda^2\sqrt{\lambda^2-\bar v^2}\\
\lambda^6\bar E_0
\end{array}
\right],
\end{align}
and $\boldsymbol \eta^{(4)}$,  $\boldsymbol \eta^{(5)}$, $\boldsymbol \eta^{(6)}$ are the respective complex conjugates of these three vectors.

The boundary conditions  \eqref{exp-BC} on the surfaces $x_2=\pm h$, in  terms of the generalised traction vector $\mathbf S$ require
\begin{equation}
\label{traction-BC}
\mathbf S\left(\pm kh\right)=\mathbf 0.
\end{equation}

These boundary conditions constitute six homogeneous equations that are conveniently  represented as a $6 \times 6$ matrix equation. For a non-trivial solution the determinant of the matrix must vanish. The resulting equation can be factorised to give two independent equations, which identify the configurations in which antisymmetric and symmetric propagating waves may occur  \cite{Su18, DorfOg19}. 
Specifically, for subsonic waves ($v<\lambda\sqrt{ \mu/\rho}$, so that $\bar v^2 < \lambda^2$), we find the following explicit \textit{dispersion equations},
\begin{align}
&\frac{ ( \lambda^6 - \lambda^4 \bar v^2 +1)^2 - \lambda^8 (\lambda^6- \lambda^4 \bar v^2 -1) \bar E_0^2 }{4\lambda^2 \sqrt{ \lambda^2 - \bar v^2}} \qquad\nonumber\\[4pt]
& \qquad\qquad = \left[ \frac{\tanh (kH\sqrt{\lambda^2 - \bar v^2} )}{\tanh (kH\lambda^{-2})} \right] ^{\pm 1} \label{dispersion} ,
\end{align} 
where the exponents $\pm 1$ correspond to antisymmetric and symmetric modes, respectively. 

To evaluate the response in the short-wave/thick-plate and long-wave/thin-plate limits, we note that $kH=2 \pi H/\mathcal L$, where $\mathcal L$ denotes the wavelength.  Therefore, for short wavelengths/thick plates (Rayleigh surface waves), $kH\rightarrow \infty$ and equation \eqref{dispersion} specialises  to 
\begin{equation}
\label{RayleighSpeed}
\frac{ ( \lambda^6 - \lambda^4 \bar v^2 +1)^2 - \lambda^8 (\lambda^6- \lambda^4 \bar v^2 -1) \bar E_0^2 }{4\lambda^2 \sqrt{ \lambda^2 - \bar v^2}}=1.
\end{equation} 
In the long wavelength/thin plate limit, $kH\rightarrow 0$ and the antisymmetric mode simplifies to
\begin{equation}
\label{long-waves-a} 
\lambda^8  \bar E_0^2  -\lambda^4\left(\lambda^2 - \bar v^2\right) +1=0,
\end{equation} 
while symmetric incremental modes occur when
\begin{equation}
\label{long-waves-s} 
\lambda^8  \bar E_0^2  -\lambda^4\left(\lambda^2 - \bar v^2\right) -3=0.
\end{equation} 
As expected, the above equations recover the purely elastic case \cite{OgRox93} when $\bar E_0=0$ and the static electro-elastic case \cite{Su18} when $\bar v=0$.

When $\bar v^2 > \lambda^2$, using $\tanh (\mathrm i x) = \mathrm i \tan (x)$, we obtain 
\begin{align}
&\frac{ ( \lambda^6 - \lambda^4 \bar v^2 +1)^2 - \lambda^8 (\lambda^6- \lambda^4 \bar v^2 -1) \bar E_0^2 }{4\lambda^2 \sqrt{\bar v^2- \lambda^2}} \nonumber\\
&\qquad \qquad = \mp\left[ \frac{\tan (kH\sqrt{\bar v^2-\lambda^2 } )}{\tanh (kH \lambda^{-2})} \right] ^{\pm 1} \label{dispersion-2} ,
\end{align} 
where the upper and lower signs correspond to antisymmetric and symmetric modes, respectively. 
When $kH \ll 1$, Eq. \eqref{dispersion-2} reduces to \eqref{long-waves-a} for antisymmetric modes and to \eqref{long-waves-s} for symmetric modes.
When $kH \rightarrow \infty$, the limit is indeterminate, as the limit of $\tan (kH \sqrt{\bar v ^2 - \lambda^2})$ is then undefined.
This is  expected because supersonic Rayleigh waves do not exist here;
as seen in Figure \ref{Rayleigh}, $\bar v^2$ is always less than $\lambda^2$ for thick plates.


\section{Numerical results and discussion}
\label{results}



\subsection{Neo-Hookean dielectric plate}

\begin{figure}[h!]
\centering
\includegraphics[width=0.98\columnwidth]{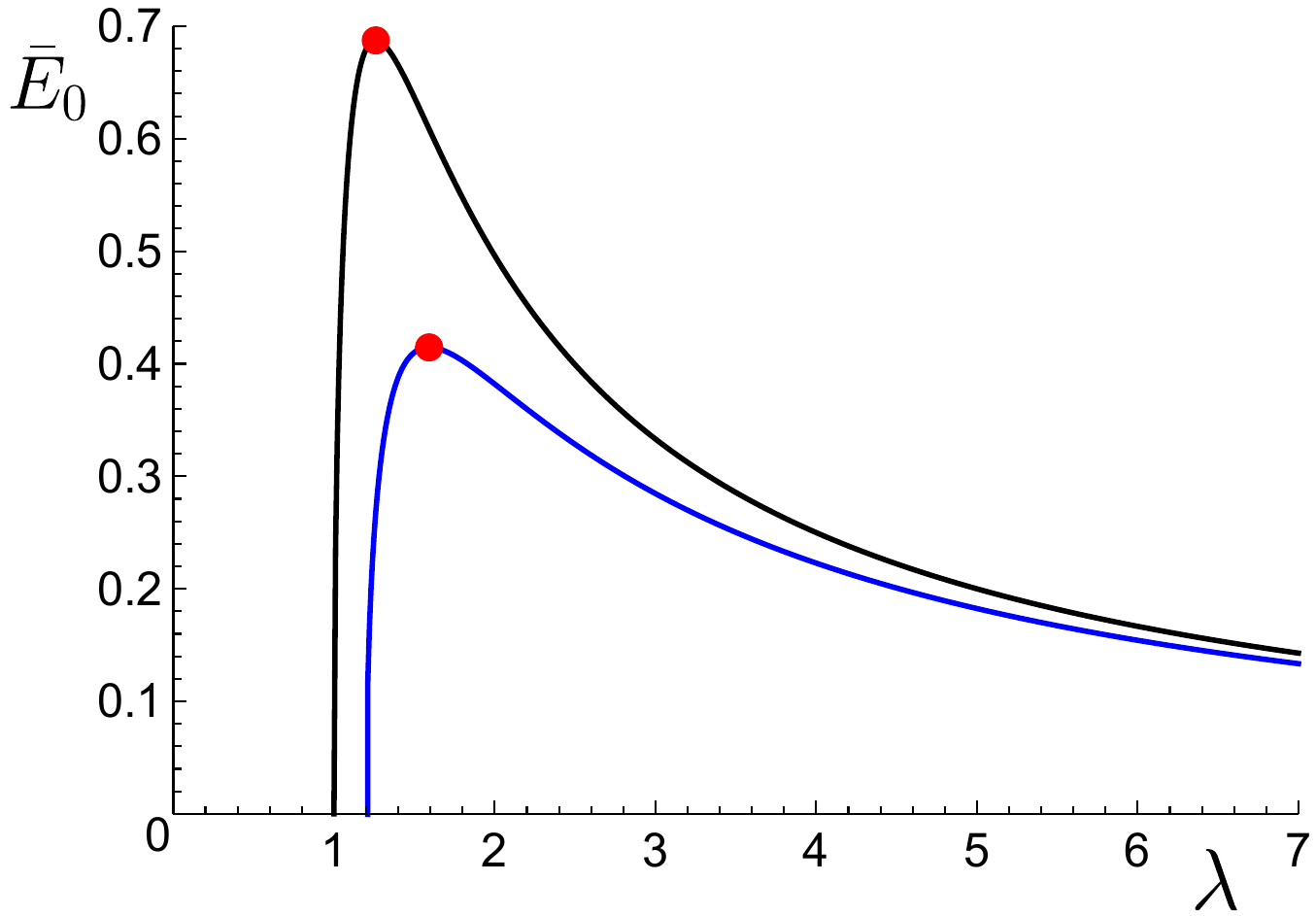}
\caption{The loading curve $\bar E_0$ against the in-plane stretch $\lambda$   for a neo-Hookean electroelastic plate with prestress $\bar T=0$ (upper curve) and $\bar T = 0.8$ (lower curve). The  field $\bar E_0$  increases initially with increasing values of $\lambda$, obtains its maximum $\bar E_{\mathrm{max}}$  (marked by {\large$\bullet$}) and decreases afterwards.  As the pre-stress increases, $\bar E_{\mathrm{max}}$ decreases, see \cite {Su18} for a detailed discussion.}
\label{LoadingCurve}
\end{figure}

On specialising to the constitutive model \eqref{neo-Hookean} and using the boundary condition $\tau_{22}=0$,  the in-plane nominal stress $T=\tau_{11}/\lambda=\tau_{33}/\lambda$ is obtained as
\begin{equation}
\label{nominal}
T=\mu\left(\lambda-\lambda^{-5}\right)-\varepsilon \lambda^3 E^2_{\mathrm L2},
\end{equation}
or, equivalently
\begin{equation}
\label{E_0}
\bar E_0=\sqrt{\lambda^{-2}-\lambda^{-8}-\lambda^{-3}\bar T},
\end{equation}
where $\bar T=T/\mu$ is a dimensionless measure  of the nominal stress. 
Relation \eqref{E_0} is illustrated by the loading curve $\bar E_0$ against the in-plane stretch $\lambda$,  with pre-stresses $\bar T=0$  and $\bar T = 0.8$ in  Figure \ref{LoadingCurve}. In the absence of pre-stress, the maximum value, $\bar E_{\mathrm{max}}=\sqrt{3}/2^{4/3} \simeq 0.69$, that the plate can support occurs at the critical stretch  $\lambda=2^{1/3} \simeq 1.26$, as shown by Zhao and Suo \cite{ZhSu07}. 
Loading curves for increasing amounts of prestress show a continuous reduction in the value of $\bar E_{\mathrm{max}}$, see for example \cite{ZhSu07, Su18,DorfmannOgden2019}. 

\begin{figure}[t!]
\centering
\includegraphics[width=0.98\columnwidth]{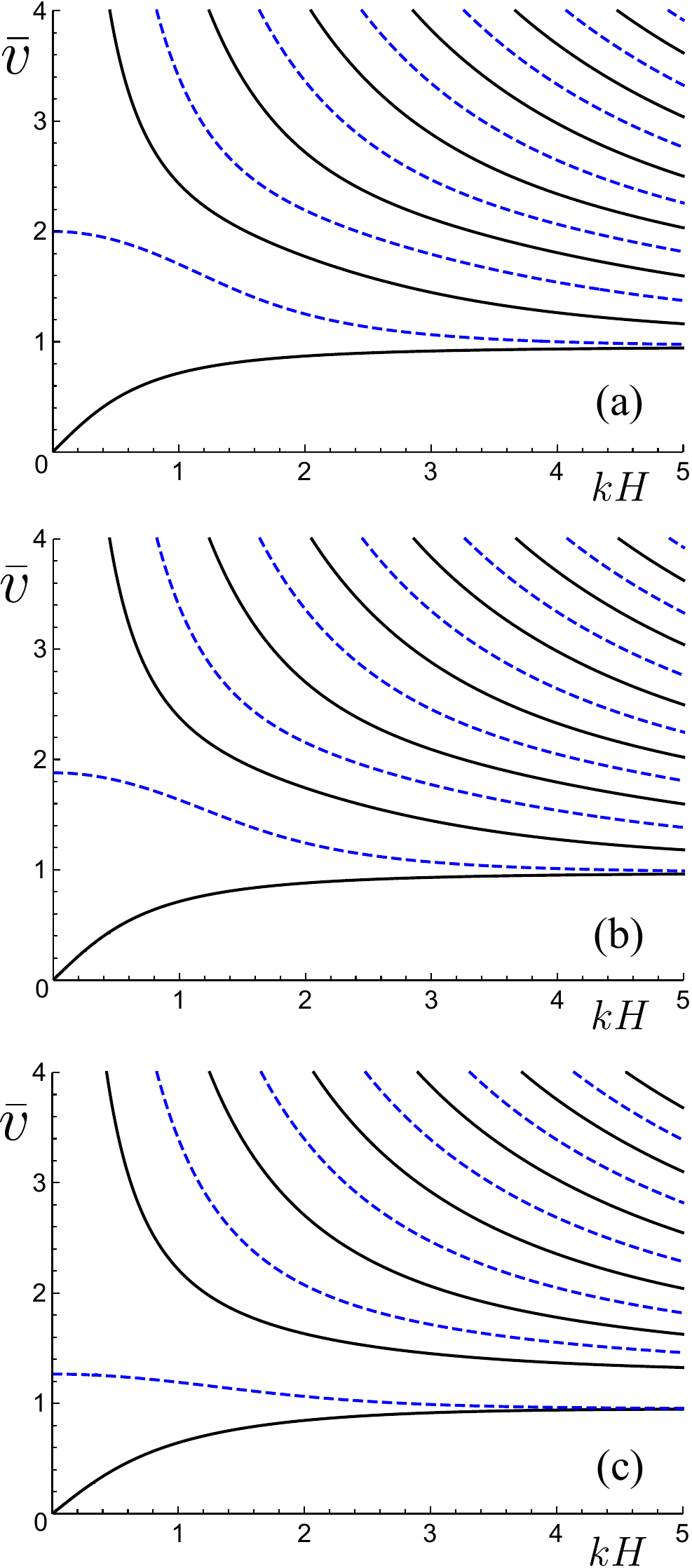}
\caption{The dimensionless wave speed $\bar v=v \sqrt{\rho/\mu}$ against $kH$ of antisymmetric and symmetric modes shown by solid and dashed curves, respectively, for a neo-Hookean electroelastic plate with $\bar T=0$. (a)  Purely elastic un-stretched case $\bar E_0=0, \lambda=1$, (b) Moderate electrical loading $\bar E_0=0.4$, $\lambda \simeq 1.03$, and (c) Maximal electrical loading $\bar E_0 = \bar E_{\mathrm {max}}=\sqrt{3}/2^{4/3}$, $\lambda=2^{1/3}$.}
\label{nHDispersion}
\end{figure}

We first consider a plate in the absence of pre-stress, i.e. we take $\bar T=0$. To evaluate the effect of an electric field on the wave speeds of symmetric and antisymmetric modes, we must consider values of $\bar E_0$ and $\lambda$ on the loading curve \eqref{E_0}. Results show that for increasing values of  $\bar E_0$ the overall trend of the curves is maintained as illustrated in Figure \ref{nHDispersion}. In particular, for $\bar E_0=0$ we recover the elastic results of an un-stretched plate \cite{OgRox93}. 

\begin{figure}[h!]
\centering
\includegraphics[width=0.98\columnwidth]{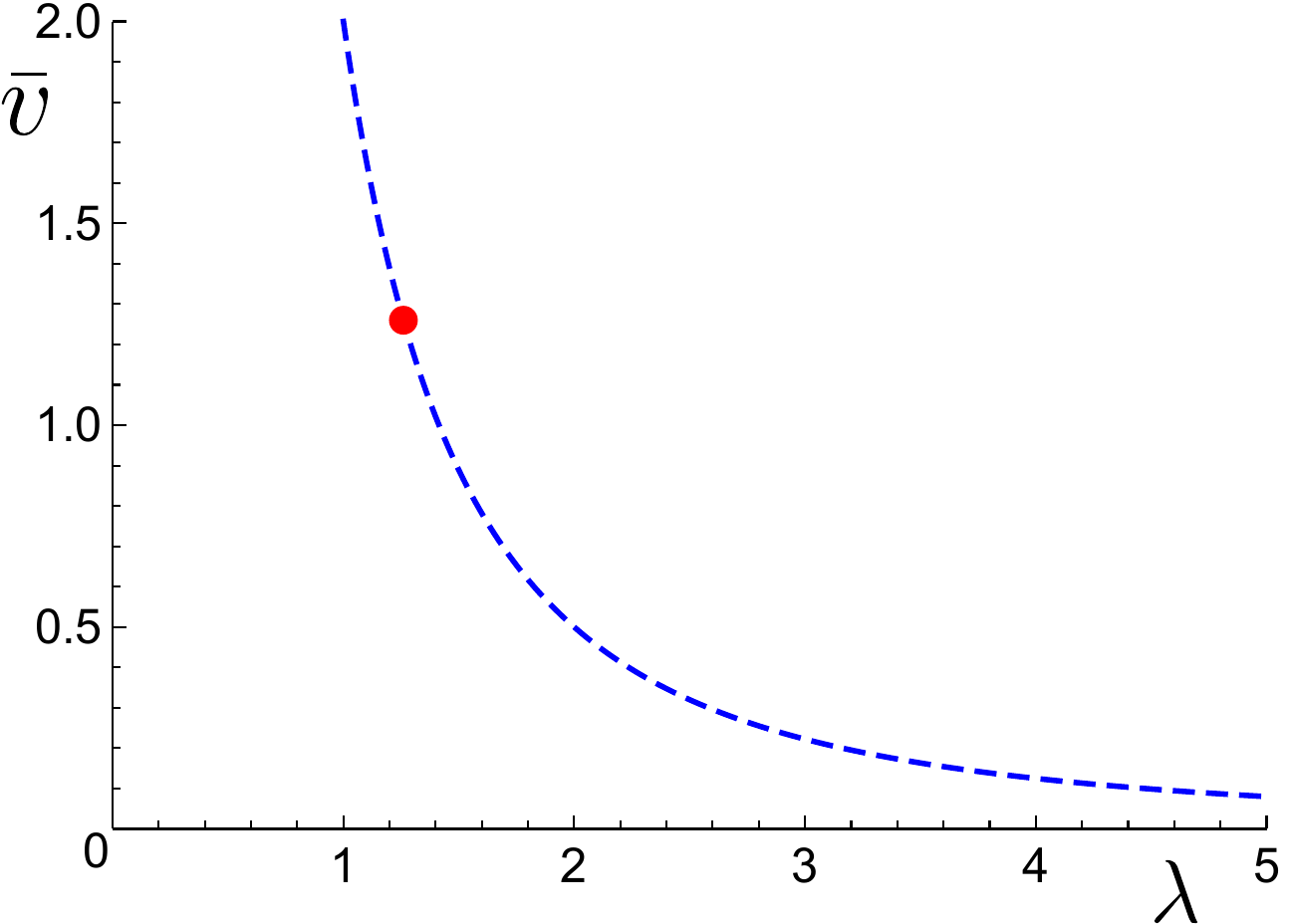}
\caption{The dimensionless velocity $\bar v=v \sqrt{\rho/\mu}$ of the fundamental symmetric mode against the stretch $\lambda$ as $kH\rightarrow 0$ (long wavelength-thin plate limit) for a neo-Hookean dielectric plate. The wave speed when  $\bar E_0 = \bar E_{\mathrm {max}}=\sqrt{3}/2^{4/3}$, $\lambda=2^{1/3}$, is marked by `{\large$\bullet$}'.}
\label{LongWaveLimit}
\end{figure}

A notable feature of the effect of $\bar E_0$ on $\bar v$   is the reduction of the speed  of the fundamental  symmetric mode in the long-wavelength/thin plate limit, i.e. the mode with finite wave speed as $kH\rightarrow 0$. 
We can see this effect directly by substituting Eq.~\eqref{E_0} at $\bar T=0$ into \eqref{long-waves-s}, giving $\bar v = 2 \lambda^{-2}$. 
The reduction in $\bar v$ of the fundamental symmetric mode with increasing values of the stretch $\lambda$ is  shown in Figure \ref{LongWaveLimit}. 

\begin{figure}[h!]
\centering
\includegraphics[width=0.98\columnwidth]{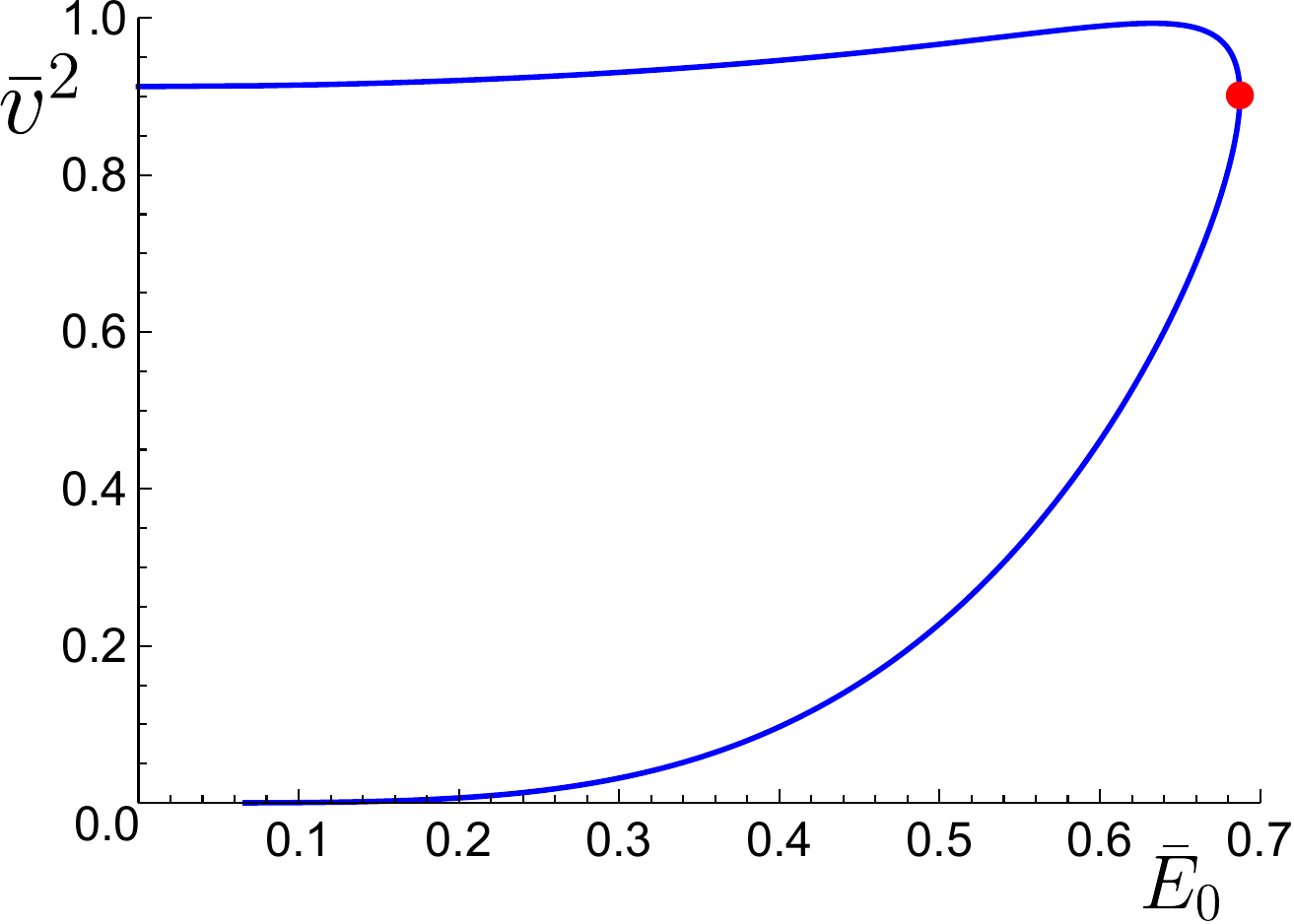}
\caption{The dimensionless squared wave speed $\bar v^2 = \rho v^2/\mu$ against $\bar E_0$  for a  neo-Hookean electroelastic plate with $\bar T=0$ and $kH\rightarrow \infty$ (short wavelength/thick plate limit).  The wave speed $\bar v^2$ when $\bar E_0 = \bar E_{\mathrm{max}}= \sqrt{3}/2^{4/3}$ is again marked by `{\large$\bullet$}'.}
\label{Rayleigh}
\end{figure}

The corresponding Rayleigh wave speed is obtained from \eqref {RayleighSpeed}  using \eqref{E_0} to express the stretch $\lambda$ in terms of the dimensionless field $\bar E_0$. This is illustrated in Figure \ref{Rayleigh} where  $\bar E_{\mathrm{max}}$ is again indicated by `{\large$\bullet$}'.   For $\bar E_0=0$, we recover Lord Rayleigh's result \cite{Rayleigh85} of the purely elastic surface wave, $\bar v^2 = 0.9126$. As $\bar E_0$ increases with increasing $\lambda$, the Rayleigh wave speed increases as well until it reaches its maximum at $\bar E_0 \simeq 0.6329$, $\lambda \simeq 1.1251$, before $\bar E_{\mathrm{max}}$ is reached. The speed then begins to decrease and past $\bar E_{\mathrm{max}}$, when the electric field decreases at increased stretch, it decreases rapidly to zero. 

\subsection{Pre-stressed dielectric plate}

We now investigate the effect of \textit{pre-stress} on wave propagation in a neo-Hookean dielectric plate. We use a pre-stress of $\bar T = 0.8$ as a representative example. 
In that case the corresponding pre-stretch is $\lambda \simeq 1.2$ in the absence of an electric field, and the loading curve reaches its maximum at $\bar E_0 = \bar E_{\mathrm{max}} \simeq 0.4148$, $\lambda \simeq 1.5916$, as seen in Figure \ref{LoadingCurve}.

\begin{figure}[h!]
\centering
\includegraphics[width=0.98\columnwidth]{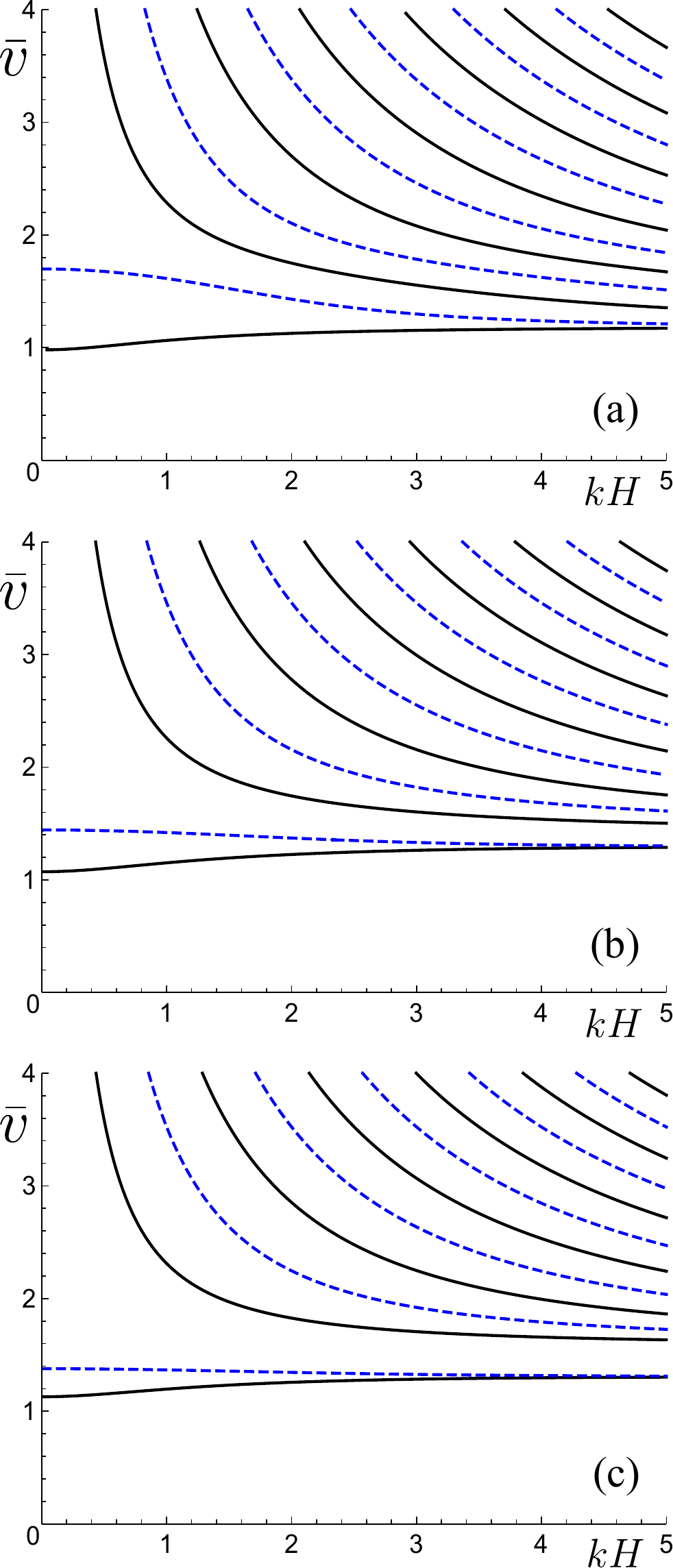}
\caption{The dimensionless wave speed $\bar v = v \sqrt{ \rho/ \mu}$ against $kH$ of antisymmetric and symmetric modes for a pre-stressed neo-Hookean electroelastic plate, with $\bar T = 0.8$. (a) The elastic case with $\bar E_0=0, \lambda\simeq 1.2$, (b) Moderate electric field $\bar E_0 = 0.4, \lambda \simeq 1.4387 $, and (c) The maximal electric field $\bar E_0 \simeq 0.4148, \lambda \simeq 1.5916$.}
\label{nHDispersion-prestretch}
\end{figure}

We again consider values of $\bar E_0$ and $\lambda$ along the loading curve \eqref{E_0}, and plot the dispersion curves  \eqref{dispersion} when $\bar T =0.8$  in Figure \ref{nHDispersion-prestretch}. When $\bar E_0=0$, the dispersion curves recover the corresponding curves in the elastic case (see, for example, \cite{OgRox93} for similar plots at different values of pre-stress). As before, as $\bar E_0$ increases towards $\bar E_{\mathrm{max}}$, the overall trend of the $\bar E_0=0$ curve is maintained. The symmetric long-wave limit decreases and the thick-plate limit increases as $\bar E_0$ is increased towards $\bar E_{\mathrm{max}}$, following similar trends to the case without pre-stress (Figs. \ref{LongWaveLimit}, \ref{Rayleigh}).

\begin{figure}[h!]
\centering
\includegraphics[width=0.98\columnwidth]{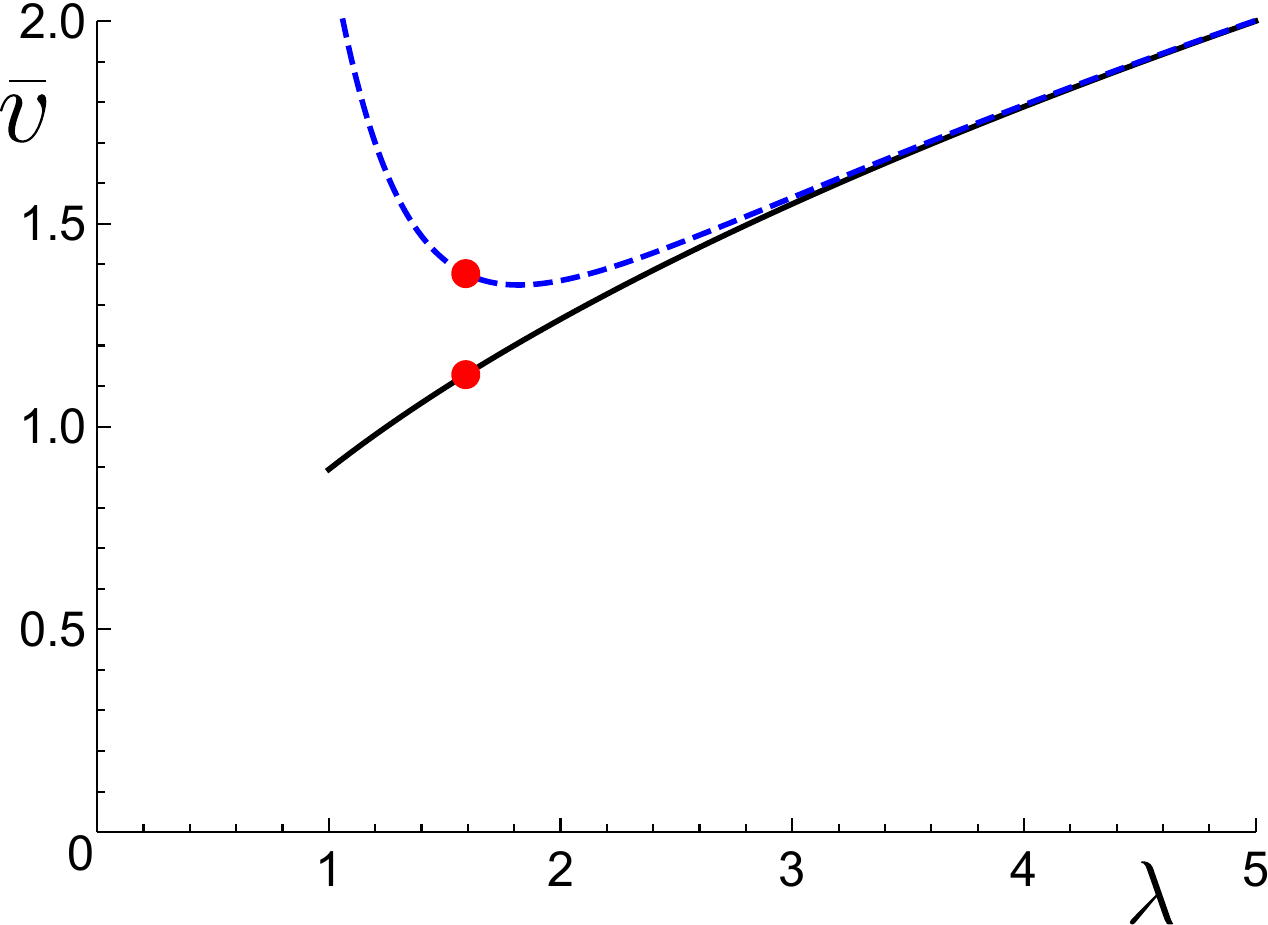}
\caption{The dimensionless wave speed $\bar v$ in the long-wave/thin-plate limit for antisymmetric (solid) and symmetric (dashed) modes in a pre-stressed electroelastic plate with $\bar T =0.8$. The wave speeds when $\bar E_0 \simeq 0.4148, \lambda \simeq 1.5916$ for both modes are marked by `{\large$\bullet$}'.}
\label{LongWaveLimit-prestretch}
\end{figure}

The major difference between the pre-stressed and non pre-stressed cases is the behaviour in  the long wavelength-thin plate regime: once a pre-stress is introduced, the speed of  the antisymmetric modes in this limit becomes non-zero. 
As the stretch is increased, the speed of the fundamental antisymmetric mode in that limit also increases (Figure \ref{LongWaveLimit-prestretch}), which can be seen explicitly by substituting the loading curve \eqref{E_0} into Eq. \eqref{long-waves-a}, giving $\bar v = \sqrt{\lambda \bar T}$, a monotonically increasing function of $\lambda$ for any positive $\bar T$.

By substituting \eqref{E_0} into \eqref{long-waves-s} we see that the  limit  for the symmetric modes becomes 
\begin{equation}
\bar v = \sqrt{4 \lambda^{-4} + \lambda \bar T}, \label{sym-pre}
\end{equation} 
in the pre-stressed case. 
As $\lambda$ increases beyond $\lambda \simeq 1.5916$, the $\lambda \bar T$ term dominates and the speed in the long wavelength-thin plate limit begins to increase, unlike in the case without pre-stress where it decreases monotonically (Fig.~\ref{LongWaveLimit}). 
As a result, for large $\lambda$ the symmetric and antisymmetric thin-plate/long wavelength limits converge to the same value ($\bar v = \sqrt{\lambda \bar T}$), as seen in Figure \ref{LongWaveLimit-prestretch}.

The thick-plate/short-wave limit also deviates from the trend in the case with no pre-stress. Initially, the wave speed $\bar v^2$ increases as the electric field increases, until it reaches the maximum $\bar E_0 \simeq 0.4148$, as before. At this point, the curve changes suddenly and the speed begins to increase, as the electric field decreases with increased stretch. At lower values of pre-stress (not shown here), the wave speed initially follows a similar trend as in Figure \ref{Rayleigh}, and begins to decrease, after which it begins to increase, as in the $\bar T = 0.8$ case. 

\begin{figure}[h!]
\centering
\includegraphics[width=\columnwidth]{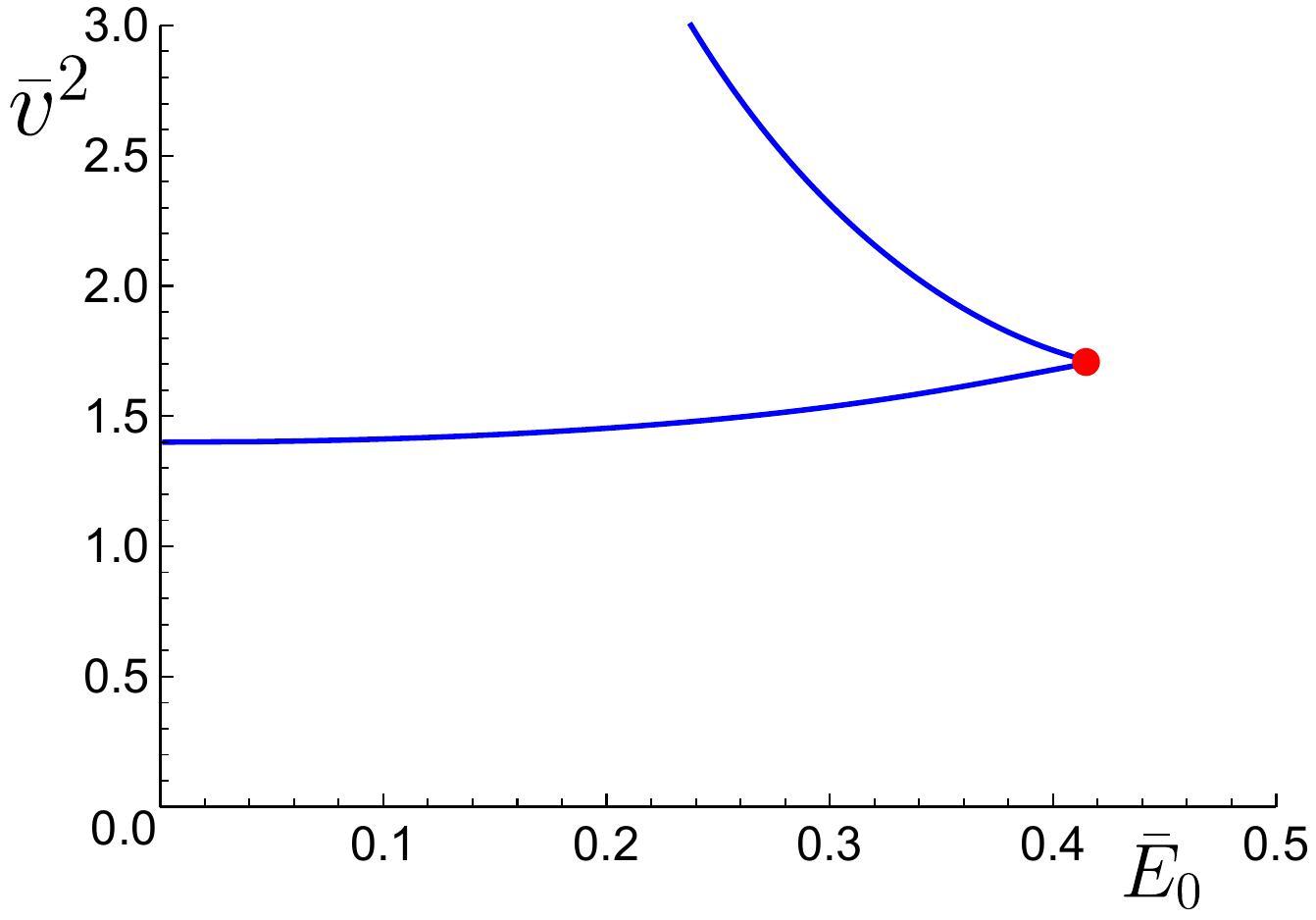}
\caption{The dimensionless wave speed $\bar v^2$ in the short-wave/thick-plate limit in a pre-stressed electroelastic plate with $\bar T =0.8$. The value of $\bar v^2$ when $\bar E_0 \simeq 0.4148, \lambda \simeq 1.5916$ is marked by `{\large$\bullet$}'.}
\label{Rayleigh-prestretch}
\end{figure}

\subsection{Gent dielectric plate}

In this section we investigate wave propagation characteristics in a plate made of an electro-elastic, \textit{strain-stiffening} Gent dielectric, with energy function 
\begin{equation}
\label{Gent}
\Omega_{\mathrm G}=-\frac{\mu J_m}{2}\log\left[1-\frac{(I_1-3)}{J_m}\right]-\frac{\varepsilon}{2}I_5,
\end{equation}
where the invariants $I_1,I_5$ are given in \eqref{I1I5again} and $J_m$ is a  material  constant, capturing strain-stiffening effects at large stretches. Following the work of Dorfmann and Ogden \cite{DorfmannOgden2014, DorfmannOgden2019} we use the specific value  $J_m=97.2$, which is typical for soft rubber. 

The in-plane nominal stress is now connected to the stretch $\lambda$ and the Lagrangian electric field  by
\begin{equation}
\label{GentNominal}
T=\mu\frac{\lambda-\lambda^{-5}}{1-(2 \lambda^2+\lambda^{-4}-3)/J_m}-\varepsilon \lambda^3 E_{\mathrm L 2}^2,
\end{equation}
leading to  the following non-dimensional form of the loading curve 
\begin{equation}
\label{GentE0}
\bar E_0=\sqrt{\frac{\lambda^{-2}-\lambda^{-8}}{1-(2 \lambda^2+\lambda^{-4}-3)/J_m}-\lambda^{-3} \bar T}.
\end{equation}

\begin{figure}[h!]
\centering
\includegraphics[width=\columnwidth]{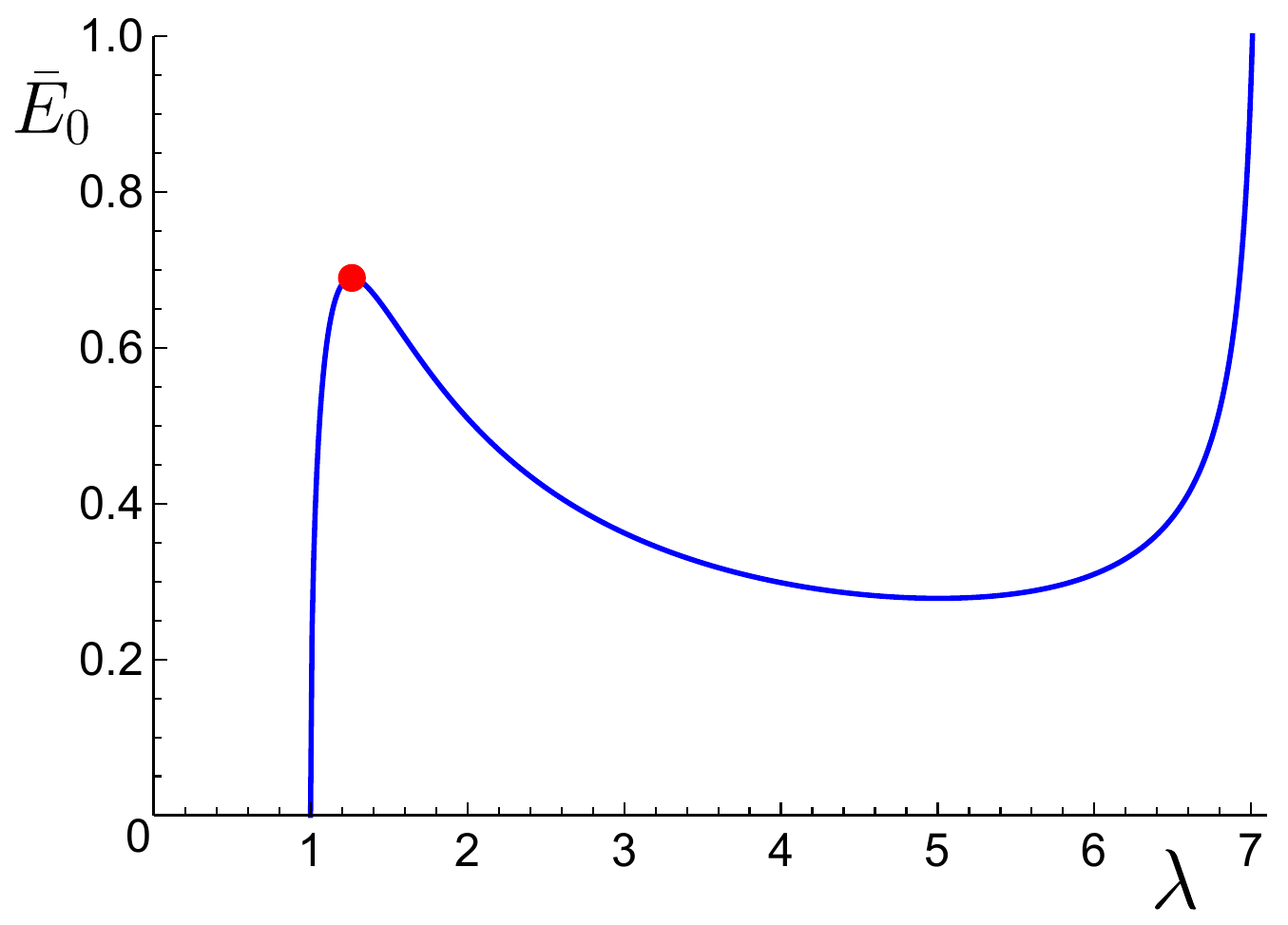}
\caption{The loading curve $\bar E_0$ against the in-plane stretch $\lambda$ for a Gent electro-elastic plate with $J_m=97.2$ and $\bar T=0$.}
\label{GentLoadingCurve}
\end{figure}

Figure \ref{GentLoadingCurve} illustrates relation \eqref{GentE0} for the special case when $\bar T=0$, highlighting the snap-through instability. The field $\bar E_0$  increases with increasing $\lambda$ and reaches a local maximum, $\bar E_{\mathrm{max}} \simeq 0.689$ when $\lambda \simeq 1.264$. The stretch then increases with constant electric field until it reaches $\lambda \simeq 6.926$, after which the stretch and electric field increase together, i.e. the snap-through instability, see \cite {Su18,DorfmannOgden2019} for detailed discussions. 

The explicit expression of the Stroh matrix $\mathbf N$ for  the Gent dielectric is obtained by following the steps identified in Section \ref{solution} with the derivatives in \eqref{coefficients} evaluated using the energy function \eqref{Gent}. Then, the solution of \eqref{eigen-problem} gives the eigenvalues \begin{align}
p_1 =&1,\label{GentValues}\\
p_{2,3} =& \frac{1}{2} \sqrt{\left(1+ \kappa \right)^2 +2(\lambda^4-\lambda^{-2})^2\frac{\bar \Omega_{11}}{\bar \Omega_1}} \notag \\& \pm \frac{1}{2}\sqrt{\left(1 - \kappa \right)^2 +2(\lambda^4-\lambda^{-2})^2\frac{\bar \Omega_{11}}{\bar \Omega_1}}, \notag
\end{align} with
\begin{equation}
\label{MoreGentValues}
p_4=-p_1,\quad p_5=-p_2,\quad p_6=-p_3,
\end{equation}  where \begin{equation}
\kappa = \sqrt{\lambda^6-\frac{\lambda^4 \bar v^2}{2 \bar \Omega _1}}, \end{equation} and \begin{align}
\bar \Omega_1&=\frac{1}{2}\frac{J_m}{3+J_m-I_1}, & \bar \Omega_{11}&=\frac{1}{2}\frac{J_m}{(3+J_m-I_1)^2},
\end{align} are the (non-dimensional) derivatives of \eqref{Gent} with respect to $I_1$.
In addition, the solution of \eqref{eigen-problem} gives the corresponding eigenvectors, which are quite lengthy and therefore not listed here.

For antisymmetric wave modes and $\bar v^2/(2 \bar \Omega_1) < \lambda^2$ we find the following dispersion equation
 \begin{strip}
\begin{multline}
 \kappa (p_2^2-p_3^2)\lambda^8 \bar E_0^2 \tanh{(\lambda^{-2}kH)} + p_3 (p_3^2 +1) \left[ 2 \bar{\Omega}_1  (\lambda^6 +p_2^2) - \lambda^4 \bar v^2 \right] \tanh{(\lambda^{-2}p_2kH)}  \\ -  p_2 (p_2^2+1) \left[ 2 \bar{\Omega}_1  (\lambda^6 +p_3^2) - \lambda^4 \bar v^2 \right] \tanh{(\lambda^{-2}p_3kH)} =0, \label{ASGent}
\end{multline} \end{strip}
\\[-14pt]
\noindent where the in-plane stretch $\lambda$ and the field $\bar E_0$ are connected by \eqref{GentE0}. 
 The corresponding equation for symmetric modes is obtained from \eqref{ASGent} by replacing the hyperbolic function $\tanh$  with $\coth$. 
 Note that for $\bar v^2/(2\bar\Omega_1) > \lambda^2$ the eigenvalue $p_2$ is imaginary and, therefore, to obtain real solutions the dispersion equation for antisymmetric and symmetric modes must be multiplied by the imaginary unit $\mathrm i$. 

Again the dispersion relation \eqref{ASGent} recovers the electrostatic case when $\bar v =0$ and the elastic case when $\bar E_0=0$.

\begin{figure}[h!]
\centering
\includegraphics[width=\columnwidth]{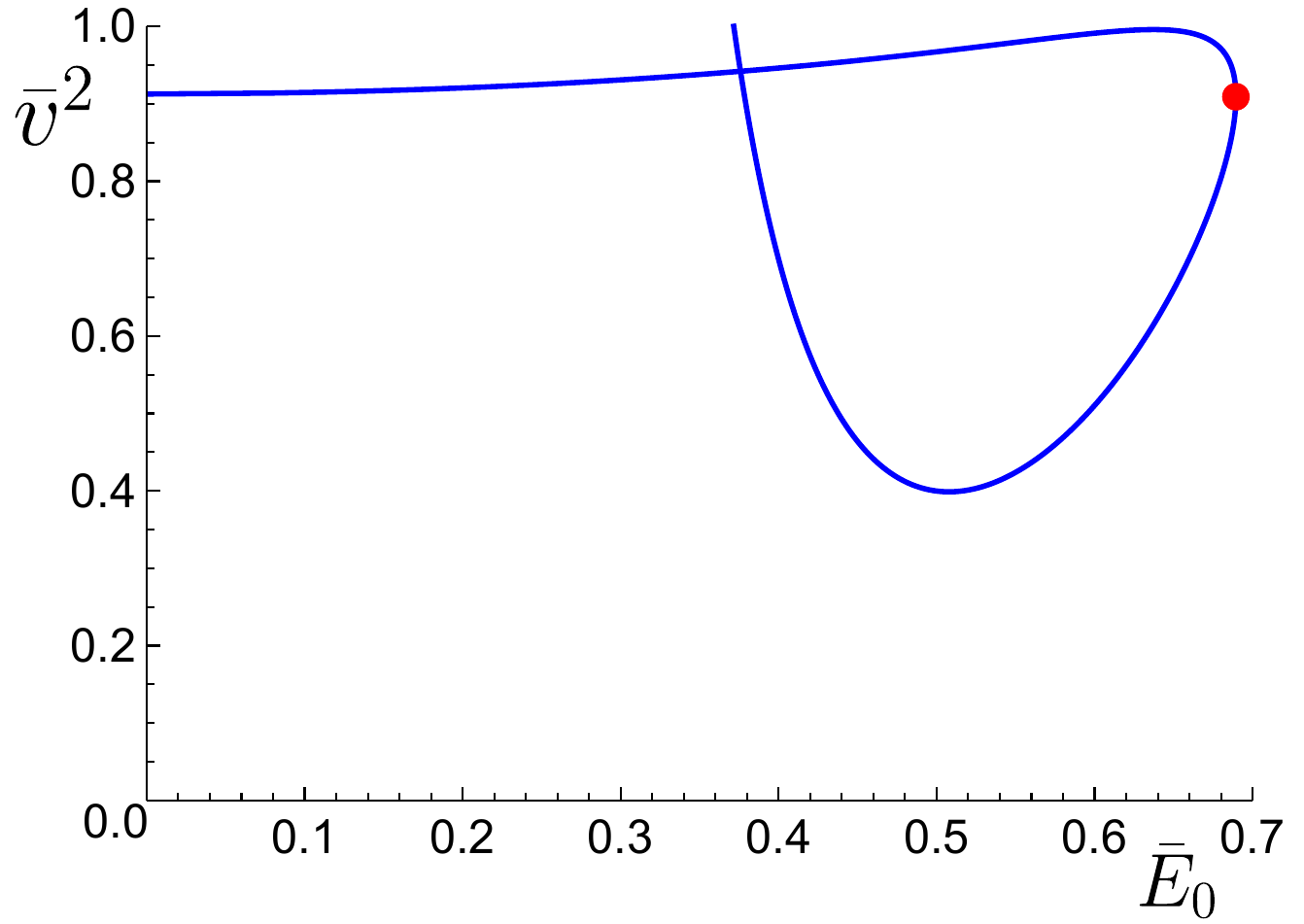}
\caption{The dimensionless wave speed $\bar v^2$ against  $\bar E_0$   for a Gent electro-elastic plate with  $\bar T=0$ in the short wavelength-thick plate limit. 
The value of  $\bar v^2$ corresponding to  $\bar E_0 = \bar E_{\mathrm{max}}= 0.689$   is denoted by `{\large$\bullet$}'.}
\label{GentRayleigh}
\end{figure}

When $kH \gg 1$ (Rayleigh waves), Equation \eqref{ASGent} and its symmetric counterpart tend to
\begin{align}
&\lambda^8 \kappa (p_2+ p_3) \bar E_0^2 +\lambda^4 \bar v^2 (p_2^2 + p_3^2 +1+ \kappa) \notag \\
& -2 \left[ \lambda^6 (p_2 +p_3)^2 +(1-\kappa)(\lambda^6-\kappa) \right] \bar \Omega_1 = 0.\label{GentShortWave}
\end{align}
The Rayleigh wave speed $\bar v$ in \eqref{GentShortWave} is evaluated using  \eqref{GentE0} and depicted in Figure \ref{GentRayleigh} against the field $\bar E_0$. 
For the purely elastic case with no pre-stress ($\bar E_0=0$, $\lambda=1$) we again recover the Rayleigh wave speed of a linear incompressible solid \cite{Rayleigh85}, $\bar v^2=0.9126$ (see \cite{DeSc04, DeOg05} for Rayleigh waves in a deformed, purely elastic, Gent material). Initially, the wave speed follows a similar trend as in the neo-Hookean case (Figure \ref{Rayleigh}), but shortly after the snap-through point ($E_0 \simeq 0.689$) the speed begins to increase substantially with decreasing electric field, due to the exponential stiffening of the material with increasing stretch after the snap-through instability.

\begin{figure}[h!]
\centering
\includegraphics[width=\columnwidth]{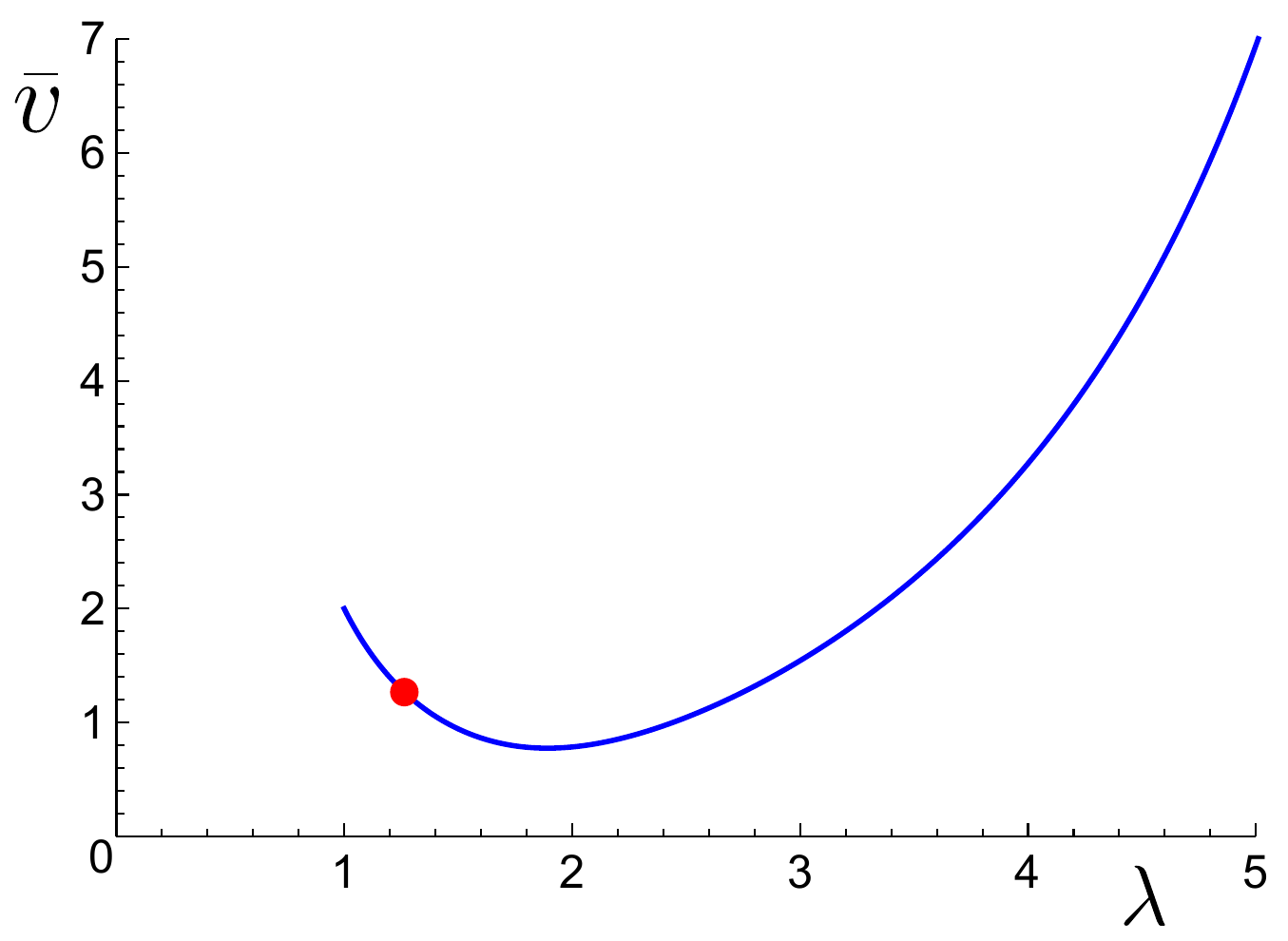}
\caption{Plots of the dimensionless velocity $\bar v$ of the fundamental symmetric mode as  $kH \rightarrow 0$ against the stretch $\lambda$. The wave speed when $\lambda = 1.264$ is marked by `{\large$\bullet$}' and occurs when the non-dimensionalised field  $\bar E_0 = \bar E_{\mathrm{max}} =  0.689$.}
\label{GentLongWaveLimit}
\end{figure}

In the long wavelength-thin plate  limit $kH \rightarrow 0$, the dispersion equation for antisymmetric incremental modes simplifies to
\begin{equation}
\label{AS-kH0}
\lambda^8 \bar E_0^2 -2 \bar \Omega_1 (\lambda^6-1) +\lambda^4 \bar v^2 =0,
\end{equation}
while the wave speed for symmetric  modes is governed by
 \begin{equation}
 \label{Sym-kH0}
\lambda^{12} \bar E_0^2 +\lambda^8 \bar v^2 -4 \bar \Omega_{11} (\lambda^6  - 1)^2 -2\lambda^4 \bar \Omega_1 (\lambda^6+3) = 0. 
\end{equation} 
It is easy to check that Eq.~\eqref{AS-kH0} gives the trivial solution $\bar v=0$ for $\bar E_0$ and $\lambda$ taken on the loading curve  \eqref{GentE0}, while Eq.~\eqref{Sym-kH0} simplifies to 
   \begin{equation}
 \label{Sym-kH0-v2}
 8\lambda^4 \bar \Omega_1 + 4 \bar \Omega_{11} (\lambda^6  - 1)^2 - \lambda^8 \bar v^2 = 0. 
\end{equation} 
The corresponding non-dimensionalised velocity $\bar v$ of the symmetric fundamental mode in the limit  $kH \rightarrow 0$ against the in-plane stretch $\lambda$ is shown in Figure \ref{GentLongWaveLimit}. As $\lambda$ increases towards the snap-through point ($\lambda \simeq 1.264$), the speed of the symmetric fundamental mode  decreases, as in the neo-Hookean case (Figure \ref{LongWaveLimit}). When $\lambda \simeq 1.9$, the speed begins to increase as the stretch increases, and continues to increase monotonically beyond the past snap-through point ($\lambda \simeq 6.926$). This effect is due to the stiffening after the snap-through instability.

\begin{figure}[t!]
\centering
\includegraphics[width=0.98\columnwidth]{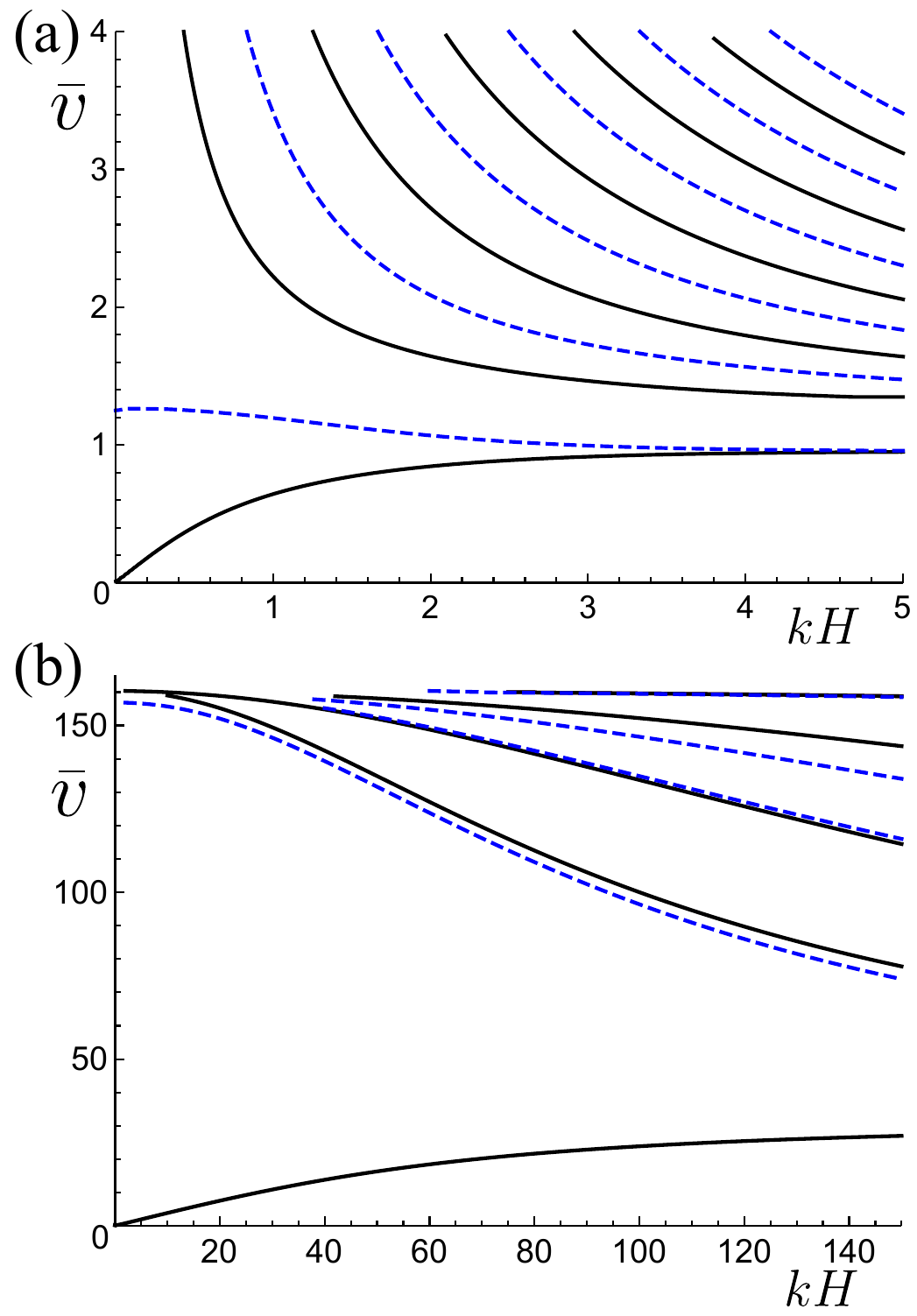}
\caption{The non-dimensionalised wave speed $\bar v=v \sqrt{\rho/\mu}$ against $kH$ of antisymmetric and symmetric modes shown by solid and dashed curves, respectively,  for a Gent electroelastic plate with $\bar T=0$. (a) The response at the snap-through point where $\bar E_{\mathrm {max}}=0.689$ and  $\lambda = 1.264$, and (b) The past snap-through point where $\bar E_{\mathrm {max}}=0.689$ and  $\lambda = 6.926$, see Figure \ref{GentLoadingCurve}.  }
\label{GentDispersion}
\end{figure}

To illustrate the effect of the snap-through instability on the wave velocity, we plot the dispersion relation \eqref{ASGent} at the snap-through point ($\bar E_0 = \bar E_{\mathrm{max}} \simeq 0.689$, $\lambda \simeq 1.264$) and the past snap-through point where $\lambda \simeq 6.926$ in Figure \ref{GentDispersion}. 
At the snap-through point ($\lambda \simeq 1.264$), the curves follow the same trend as those of the neo-Hookean energy density function, see Figure \ref{nHDispersion}c. After the snap-through has taken place ($\lambda \simeq 6.926$), the velocity increases dramatically. 
The speeds of the symmetric modes in the $kH \ll 1$ and $kH \gg 1$ regimes have both increased, but the overall trend of the fundamental modes remains the same. 
The dramatic increase in the values of the speed is due to the very large stretch at the past snap-through point, where the plate is under large strain and is much stiffened. 

The first antisymmetric and symmetric modes also converge to the short-wave limit much slower than in all other cases. This is again due to the large stretch, as the competition between $kH$ and $\lambda^{-2}$ is greater for large $\lambda$.


\section{Concluding remarks} 


In this paper, we investigated Lamb wave characteristics in the presence of an electric field generated by a potential difference between two flexible electrodes mounted on the major surfaces of a dielectric plate.
The incremental governing equations and the electrical and mechanical boundary conditions are given in Stroh form and solved numerically for neo-Hookean and Gent dielectric models. The dispersion equation is factorised to give two independent equations, which identify the configurations where antisymmetric and symmetric propagating waves may occur. Explicit expressions of the dispersion equations for the short-wavelength/thick-plate and long-wavelength/thin-plate limit are obtained.  In particular, we investigated the effects of plate thickness, of the electric field, of an in-plane pre-stress and of stretch-stiffening on the wave characteristics.


\section*{Acknowledgement}


This work is supported by a Government of Ireland Postgraduate Scholarship from the Irish Research Council (Project GOIPG/2016/712).

\end{document}